\documentstyle{article}

\def\be{\begin{equation}}
\def\ee{\end{equation}}
\def\bea{\begin{eqnarray}}
\def\eea{\end{eqnarray}}
\begin{document}
\begin{titlepage}
\vspace*{10mm}
\begin{center}
{\large \bf Two--dimensional fractional supersymmetric conformal-
and logarithmic conformal- field theories and the two point functions}
\vskip 2\baselineskip
\centerline {\bf
 Fardin Kheirandish $^{a}$ \footnote {e-mail:fardin@iasbs.ac.ir} {\rm and}
 Mohammad Khorrami $^{a,b}$ \footnote {e-mail:mamwad@iasbs.ac.ir}}
\vskip 2\baselineskip
{\it $^a$ Institute for Advanced Studies in Basic Sciences,}
\\ {\it P. O. Box 159, Zanjan 45195, Iran}\\
{\it $^b$ Institute for Studies in Theoretical physics and
Mathematics,}\\ {\it P. O. Box 5531, Tehran 19395, Iran}\\
\end{center}
\vskip 2cm
\begin{abstract}
\noindent A general two--dimensional fractional supersymmetric
conformal field theory is investigated. The structure of the
symmetries of the theory is studied. Applying the generators of
the closed subalgebra generated by $(L_{-1},L_{0},G_{-1/(F+1)})$
and $(\bar{L}_{-1},\bar{L}_{0},\bar{G}_{-1/(F+1)})$, the two point
functions of the component--fields of supermultiplets are
calculated. Then the logarithmic superconformal field theories are
investigated and the chiral and full two--point functions are obtained.
\end{abstract}
\end{titlepage}
\newpage
\section{Introduction}
2D conformally--invariant field theories have become the subject
of intense investigation in recent years, after the work of
Belavin, Polyakov, and Zamolodchikov [1]. One of the main reasons
for this, is that 2D conformal field theories describe the
critical behaviour of two dimensional statistical models [2--5].
Conformal field theory provides us with a simple and powerful
means of calculating the critical exponents, as well as, the
correlation functions of the theory at the critical point [1,6].
Another application of conformal field theories is in string
theories. Originally, string theory was formulated in flat 26--dimensional
space--time for bosonic- and flat 10-dimensional space--time for
supersymmetric-theories. It has been realized now that the central part of
string theory is a 2D conformally--invariant field theory. It is seen that
the tree--level string amplitudes may be expressed in terms of the
correlation functions of the corresponding conformal field theory on the
plane, whereas string loop amplitudes may be expressed in terms of the
correlation functions of the same conformal field theory on higher--genus
Riemann surfaces [7--10].

Supersymmtry is a $Z_{2}$ extension of the Poincar\'e algebra
[11,12]. But this can be enlarged to a superconformal algebra ([13] for
example). If the dimension of the space--time is two, there
are also fractional supersymmetric extensions of the Poincar\'e
and conformal algebra [14--17]. Fractional supersymmetry is a
$Z_{n}$ extension of the Poincar\'e algebra. In this paper, the genral
form of the two--point functions of a theory with this symmetry is obtained.

According to Gurarie [18], conformal field theories with logarithmic
correlation functions may be consistently defined. In some interesting
physical theories like polymers [19], WZNW models [20--23], percolation
[24], the Haldane-Rezayi quantum Hall state [25], and edge excitation in
fractional quantum Hall effect [26], there appear logarithmic correlation
functions. Logarithmic operators are also seen in 2D magnetohydrodynamic
turbulence [27--29], 2D turbulence [30,31] and some critical
disordered models [32,33]. Logarithmic conformal field theories
for D--dimensional case ($D>2$) has also been studied [34]. The general form
of the correlators of the 2D conformal field theories and supersymmetric
conformal field theories is investigated in [35,36] and [37], respectively.

In this paper, the general case $n=F+1$ is considered. So the components
of the superfield have grades 0, 1,..., and $F$. The complex plane
is extended by introducing two independent paragrassmann variables
$\theta$ and $\bar{\theta}$, satisfying
$\theta^{F+1}={\bar{\theta}^{F+1}}=0$. One can develop an algebra,
the fractional $n=F+1$ algebra, based on these variables and the derivatives
with respect to them [38--40]. In [41], the fractional supersymmetry has
been investigated by introducing a certain fractional
superconformal action. We don't consider any special action here.
What we do, is to use only the structure of the fractional
superconformal symmetry to obtain general restrictions on the form of
the two--point functions. The scheme of the paper is the
following. In section 2, infinitesimal superconformal
transformations are defined. In section 3, the generators of these
transformation and their algebra are investigated. In section 4,
the two--point functions of such theories are obtained. In fact,
sections 2--4 are a generalization of [42]. In section 5,
logarithmic superconformal field theories are investigated and
the chiral- and full- two--point functions of primary and quasiprimary
(logarithmic) fields are obtained. This is a generalization of [43].

\section{Infinitesimal superconformal transformations}
Consider a paragrassmann variable $\theta$, satisfying
\be
\theta^{F+1}=0,
\ee
where $F\geq 2$, is a positive integer. A function of a complex variable
$z$, and this paragrassmann variable, will be of the form
\be
f(z,\theta)=f_{0}(z)+\theta
f_{1}(z)+\theta^{2}f_{2}(z)+...+\theta^{F}f_{F}(z).
\ee
The covariant derivative is defined as [16,38,44,45]
\be
D:=\partial_{\theta}+\frac{q^F\theta^{F}}{(F)_{q^{-1}}!}\partial_{z},
\ee
where
\be
(F)_q!:=(F)_{q}(F-1)_{q}...(1)_{q}, \hspace{1 cm}
(m)_{q}=\frac{1-q^{m}}{1-q},
\ee
and
\be
\partial_{\theta}\theta=1+q\theta\partial_{\theta}.
\ee
$q$ is an $F+1$th root of unity, with the property that there exisits no
positive integer $m$ less than $F+1$ so that $q^m=1$.

An Infinitesimal transformation
\bea
z'&=&z+\sum_{k=0}^{F}\theta^{k}\omega_{k}(z),\cr
\theta'&=&\theta+\sum_{k=0}^{F}\theta^{k}\epsilon_{k}(z),
\eea
is
called superconformal if
\be
D=(D\theta')D',
\ee
where
\be
D'=\partial_{\theta'}+\frac{q^F\theta'^{F}}{(F)_{q^{-1}}!}\partial_{z'},
\ee
from these, it is found that an infinitesimal superconformal
transformation is of the form
\bea
 z'&=&z+\omega_{0}(z)+\frac{q^F\theta^F}{(F)_{q^{-1}}!}\epsilon_{0}(z),\cr
 \theta'&=&\theta+\epsilon_{0}(z)+\frac{1}{F+1}\theta\omega_{0}'(z)+
 \sum_{k=2}^{F}\theta^{k} \epsilon_{k}(z),
\eea
where $\omega_{0}'(z):=\partial_{z}\omega_{0}(z)$. It is also
seen that the following commutation relations hold
\be
{\epsilon_{i}\theta}={q\theta\epsilon_{i}},\hspace{1 cm}i\neq 1.
\ee
One can extend these naturally to functions of z and $\bar{z}$, and $\theta$
and $\bar{\theta}$ (full functions instead of chiral ones). It is sufficient
to define a covariant derivative for the pair $(\bar{z}, \bar\theta)$, the
analogue of (3), and extend the transformations (5), so that there
are similar transformations for $(\bar{z}, \bar{\theta})$ as well.
Then, defining a superconformal transformation as one satisfying
(6) and its analogue for $(\bar{z}, \bar{\theta})$, one obtains,
in addition to (8) and (9), similar expressions where $(z, \theta,
\omega_{k}, \epsilon_{k})$ are simply replaced by $(\bar{z},
\bar{\theta}, \bar{\omega}_{k}, \bar{\epsilon}_{k})$. So, the
superconformal transformations consists of two distinct class of
transformations, the holomorphic and the antiholomorphic, that do
not talk to each other.
\section{generators of superconformal field theory}
The (chiral) superfield $\phi(z,\theta)$, with the expansion,
\be
\phi(z,\theta)=\varphi_{0}(z)+\theta\varphi_{1}(z)+\theta^{2}\varphi_{2}(z)+...+\theta^{F}
\varphi_{F}(z),
\ee
is a super--primaryfield of the weight $\Delta$, if it transforms under a
superconformal transformation as
\be
\phi(z,\theta)\mapsto (D\theta')^{(F+1)\Delta}\phi(z',\theta').
\ee
One can write this as
\be
\phi(z,\theta)\mapsto[1+\tilde{T}(\omega_{0})+\tilde{S}(\epsilon_{0})+
\sum_{k=2}^{F}\tilde{H}_{k}(\epsilon_{k})]\phi(z',\theta'), \ee to
arrive at [38] \bea
\tilde{T}(\omega_{0})&=&\omega_{0}\partial_{z}+(\Delta+
\frac{\Lambda}{F+1})\omega_{0}',\cr
\tilde{S}(\epsilon_{0})&=&\epsilon_{0}\left[\delta_{\theta}+
\frac{\theta^{F}}{(F)_{q^{-1}}!}\partial_{z}\right]+
\frac{F+1}{(F)_{q^{-1}}}\Delta\epsilon_{0}'\theta^{F},\cr
\tilde{H}_{k}(\epsilon_{k})&=&\theta^{k}\epsilon_{k}\delta_{\theta}.
\eea Here $\Lambda$ and $\delta_{\theta}$, are operators
satisfying
\be
[\Lambda,\theta]=\theta,\hspace{1 cm}
[\Lambda,\delta_{\theta}]=-\delta_{\theta},
\ee
and
\be
\delta_{\theta}\theta=q^{F}\theta\delta_{\theta}+1.
\ee
One can now define the classical generators
\bea
l_{n}&:=&\tilde{T}(z^{n+1}),\cr
g_{r}&:=&\tilde{S}(z^{r+1/(F+1)}),
\eea
where, $n$ and $r+1/(F+1)$ are integers. We do not consider the generators
corresponding to $\tilde{H}$, since there is no closed subalgebra, with a
trivial central extension, containing these generators [16,46]. The quantum
generators of superconformal transformations are defined through
\bea
{[L_{n},\phi(z,\theta)]}&:=&l_{n}\phi,\cr
{[G_{r},\phi(z,\theta)]}&:=&g_{r}\phi.
\eea
One can chek that, apart from a possible central extension, these generator
satisfy the following relations.
\bea
{[L_{n}, L_{m}]}&=&(n-m)L_{n+m},\\
{[L_{n}, G_{r}]}&=&\left(\frac{n}{F+1}-r\right)G_{n+r},
\eea
and
\be
\{G_{r_{0}}G_{r_{1}}...G_{r_{F}}\}_{per}=(F+1)!
L_{\left(\sum_{k=0}^{F}r_{k}\right)}\ee where $\{...\}_{per}$,
means, the sum of products of all possible permutations of
generators $G_{r_{k}}$. This algebra has  nontrivial central
extensions, it is shown [16,46] that there is only one subalgebra
(containing $G_r$'s as well as $L_{n}$'s), the central extension
for which is trivial. This algebra is the  one generated by
$[L_{-1}, L_{0}, G_{-1/(F+1)}]$ and their antiholomorphic
counterparts $[\bar{L}_{-1}, \bar{L}_{0}, \bar{G}_{-1/(F+1)}]$.
Having the effect of $L_{n}$'s and $G_{r}$'s on the superfield, it
is not difficult to obtain their effect on the component fields.
For $L_{n}$'s, the first equation of (18) leads directly to
\be
[L_{n}, \varphi_{k}]=z^{n+1}\partial_{z}\varphi_{k}+(n+1)z^{n}
(\Delta+\frac{k}{F+1})\varphi_{k}.
\ee
This shows that the component field $\varphi_{k}$, is simply a
primary field with the weight $\Delta+\frac{k}{F+1}$. One can write (22),
also in terms of the operator--product expansion:
\be
\Re[T(\omega)\varphi_{k}(z)]\sim\frac{\partial_z\varphi_{k}(z)}{\omega-z}+
\frac{(\Delta+\frac{k}{F+1})\varphi_{k}(z)}{(\omega-z)^{2}},
\ee
where $\Re$ denotes the radial ordering and $T(z)$, is the holomorphic part
of the energy--momentum tensor:
\be
T(z)=\sum_{n}{\frac{L_{n}}{z^{n+2}}}.
\ee
For $G_{r}$'s, a little more care is needed. One defines a
$\chi$--commutator as [47]
\be
[A, B]_{\chi}:=AB-\chi BA.
\ee
It is easy to see that
\be
[A, BB']_{\chi\chi'}=[A, B]_{\chi}B'+\chi B[A, B']_{\chi'}.
\ee
Now, if we use
\be
[G, \theta]_{q}=0,
\ee
then the second equation of (18) leads to
\bea
{[G_{r},\varphi_{k}]_{q^{-k}}}&=&z^{r+\frac{1}{F+1}}q^{-k}
(k+1)_{q^{-1}}\varphi_{k+1}, \hspace{1.5 cm} 0\leq k\leq F-1,\cr
{[G_{r},\varphi_{F}]_{q^{-F}}}&=&\frac{q^{-F}}{(F)_{q^{-1}}!}\Big[
z^{r+\frac{1}{F+1}}\partial_{z}\varphi_{0}\cr
&&+(F+1)\Big(r+\frac{1}{F+1}\Big)\Delta z^{r-\frac{F}{F+1}}\varphi_{0}\Big].
\eea
This can also be written in terms of the operator--product
expansion. To do this, however, one should first define a proper
radial ordering for the supersymmetry generator and the component
fields. Defining
\be
\Re[S(w)\phi_k(z)]:=\cases{S(w)\phi_k(z),&$|w|>|z|$\cr
                          q^{-k}\phi_k(z)S(w),&$|w|<|z|$\cr}
\ee
where
\be
S(z):=\sum_r{{G_r}\over{z^{r+\frac{F+2}{F+1}}}},
\ee
one arrives at
\bea
{\Re[S(\omega)\varphi_{k}(z)]}&\sim&q^{-k}(k+1)_{q^{-1}}
\frac{\varphi_{k+1}}{\omega-z}, \hspace{1.5 cm} 0\leq k\leq F-1,\cr
{\Re[S(\omega)\varphi_{F}(z)]}&\sim&\frac{q^{-F}}{(F)_{q^{-1}}!}\left[
\frac{\partial_{z}\varphi_{0}}{\omega-z}+
\frac{(F+1)\Delta\varphi_{0}}{(\omega-z)^{2}}\right].
\eea
What we really use to restrict the correlation functions is that part of
the algebra the central extension of which is trivial, that is, the algebra
generated by  $[L_{-1}, L_{0}, G_{-1/(F+1)}]$ and their  antiholomorphic
counterparts $[\bar{L}_{-1}, \bar{L}_{0},\bar{G}_{-1/(F+1)}]$.
\section{Two--point functions}
The two--point functions should be invariant under the action of
the subalgebra generated by $L_{-1}$,$L_{0}$, and $G_{-1/(F+1)}$.
This means
\bea \langle0|[L_{-1},\phi_k\phi'_{k'}]|0\rangle&=&0,\\
\langle0|[L_0,\phi_k\phi'_{k'}]|0\rangle&=&0,\\
\langle0|[G_{-1/(F+1)},\phi_{k}\phi'_{k'}]_{q^{-k-k'}}|0\rangle&=&0.
\eea
We have used the shorthand notation $\phi_k=\phi_k(z)$ and
$\phi'_{k'}=\phi'_{k'}(z')$. $\phi$ and $\phi'$ are primary superfields of
weight $\Delta$ and $\Delta'$, respectively. Equations (32) and (33) imply
\be
\langle\phi_k\phi'_{k'}\rangle ={{A_{k,k'}}\over {(z-z')^{\Delta
+\Delta'+(k+k')/(F+1)}}}.
\ee
This is simply due to the fact that $\phi_k$ and $\phi'_{k'}$ are primary
fields of the weight $\Delta +k/(F+1)$ and $\Delta' +k'/(F+1)$,
respectively. Note that it is not required that these weights be equal to
each other, since we have not included $L_1$ in the subalgebra.

Relation (34) relates $A_{k_1,k'_1}$ with $A_{k_2,k'_2}$, if
\be
k_1+k'_1-(k_2+k'_2)=0,\qquad\hbox{mod }F+1.
\ee
Therefore, there remains $F+1$ independent constants in the $(F+1)^{2}$
correlation functions. So there are correlation functions of grade 0, 1,
2, $\cdots$, and $F$. We have the following relations between the
constants $A_{k,k'}$'s:
\bea
A_{k+1,k'}&=&-q^{-k'}\frac{1-q^{-(k'+1)}}{1-q^{-(k+1)}}A_{k,k'+1},
\qquad 0\leq k,k'\leq F-1,\cr
A_{k+1,F}&=&\frac{-q(\Delta+\Delta'+\frac{F+k}{F+1})}
{(F)_{q^{-1}}!(k+1)_{q^{-1}}}A_{k,0},\qquad 0\leq k\leq F-1,\cr
A_{F,k'+1}&=&\frac{q^{k'}(\Delta+\Delta'+\frac{F+k'}{F+1})}
{(F)_{q^{-1}}!(k'+1)_{q^{-1}}}A_{0,k'},\cr
A_{F,0}&=&q^FA_{0,F}.
\eea
Using these, one can write (35) as
\be\label{chiral}
\langle\phi_k\phi'_{k'}\rangle =A_{k+k'}f_{k,k'}(z-z'),
\ee
where
\be
A_{k+k'}:=A_{k+k',0}=A_K. \ee

So far, everything has been calculated for the chiral fields. But the
genralization of this to full fields is not difficult. Following [42] and
using exactly the same reasoning, it is seen that
\be\label{full}
\langle\varphi_{k\bar k}(z,\bar z)\varphi'_{k'\bar k'}(z',\bar z')\rangle =
A_{K\bar K}q^{-k\bar k}f_{k,k'}(z-z')\bar f_{\bar k,\bar k'}
(\bar z-\bar z').
\ee
Here $\bar f$ is the same as $f$ with $\Delta\to\bar\Delta$ and
$\Delta'\to\bar\Delta'$, and
\be
K=k+k'\qquad\hbox{mod }F+1,\qquad\qquad \bar K=\bar k+\bar k'\qquad
\hbox{mod }F+1.
\ee
\section{Logarithmic two--point functions}
Suppose that the first component--field $\varphi_0(z)$ of the chiral
superprimary field $\Phi(z,\theta)$, has a logarithmic counterpart
$\varphi_0'(z)$:
\be
[L_{n}, \varphi_0'(z)]=[z^{n+1}\partial_{z}+(n+1)z^{n}\Delta]
\varphi_0'(z)+(n+1)z^{n}\varphi_0(z).
\ee
Following [37] and [43], one can show that $\varphi_0'(z)$ is the first
component--field of a new superfield $\phi'(z,\theta)$, which is the formal
derivative of the superfield $\phi(z,\theta)$. One defines the fields
$f'_{r}(z)$ through
\be
[G_{r}, \varphi_0'(z)]=:z^{r+1/(F+1)}f'_{r}(z).
\ee
Then, by a reasoning similar to that presented in [37] and [43], that is
acting on both sides by $L_m$ and using the genralized Jacobi identity, it
is seen that all $f_r$'s are the same. Denoting this filed by $\varphi_1'$,
we have
\be
[G_{r}, \varphi_0'(z)]=:z^{r+1/(F+1)}\varphi_1'(z).
\ee
One can continue and construct other component--fields. It is found that
\bea
{[G_{r},\varphi'_{k}]_{q^{-k}}}&=&z^{r+\frac{1}{F+1}}q^{-k}
(k+1)_{q^{-1}}\varphi'_{k+1}, \hspace{1.5 cm} 0\leq k\leq F-1,\cr
{[G_{r},\varphi'_{F}]_{q^{-F}}}&=&\frac{q^{-F}}{(F)_{q^{-1}}!}\Big[
z^{r+\frac{1}{F+1}}\partial_{z}\varphi'_{0}\cr
&&+(F+1)\Big(r+\frac{1}{F+1}\Big)\Delta z^{r-\frac{F}{F+1}}\varphi'_{0}\cr
&&+(F+1)\Big(r+\frac{1}{F+1}\Big) z^{r-\frac{F}{F+1}}\varphi_{0}\Big],
\eea
and
\be
[L_{n}, \varphi'_{k}]=z^{n+1}\partial_{z}\varphi'_{k}+(n+1)z^{n}
(\Delta+\frac{k}{F+1})\varphi'_{k}+(n+1)z^{n}\varphi_k.
\ee
Combining the primed component--fields in the chiral superfield $\phi'$, one
can write the action of $L_n$'s and $G_r$'s on it as
\bea
{[L_{n},\phi']}&=&\left[z^{n+1}\partial_{z}+(n+1)z^{n}\left(\Delta+
\frac{\Lambda}{F+1}\right)\right]\phi'+(n+1)z^{n}\phi,\\
{[G_{r},\phi']}&=&z^{r+\frac{1}{F+1}}\left[\delta_{\theta}+
\frac{\theta^{F}}{(F)_{q^{-1}}!}\partial_{z}\right]\phi'\nonumber\\
&&+\frac{F+1}{(F)_{q^{-1}}!}\left(r+\frac{1}{F+1}\right)z^{r-\frac{F}{F+1}}
\Delta\theta^{F}\phi'\nonumber\\&&+\frac{F+1}{(F)_{q^{-1}}!}
\left(r+\frac{1}{F+1}\right)\theta^{F}\phi .
\eea
It is easy to see that these are formal derivatives of (18) with respect to
$\Delta$, provided one defines
\be
\phi'(z,\theta) =\frac{d\phi}{d\Delta}. \ee The two superfields
$\phi$ and $\phi'$, thus combine in a two--dimensional Jordanian
block of quasi--primary fields. The generalization of the above
results to an $m$--dimensional Jordanian block: \bea
[L_{n},\phi^{(i)}]&=&\left[z^{n+1}\partial_{z}+(n+1)z^{n}
\left(\Delta+\frac{\Lambda}{F+1}\right)\right]\phi^{(i)}\cr
&&+(n+1) z^{n}\phi^{(i-1)},\hspace{1 cm} 1\leq i\leq m-1,\eea and
\bea
[G_{r},\phi^{(i)}]&=&z^{r+\frac{1}{F+1}}\left[\delta_{\theta}+
\frac{\theta^{F}}{(F)_{q^{-1}}}\partial_{z}\right]\phi^{(i)}\cr
&&+\frac{F+1}{(F)_{q^{-1}}!}\left(r+\frac{1}{F+1}\right)z^{r-\frac{F}{F+1}}
\Delta\theta^{F}\phi^{(i)}\cr
&&+\frac{F+1}{(F)_{q^{-1}}!}\left(r+\frac{1}{F+1}\right)z^{r-\frac{F}{F+1}}
\theta^{F}\phi^{(i-1)},\hspace{0.1 cm} 1\leq i\leq m-1. \eea One
can regard these as formal derivatives of (18) with respect to
$\Delta$, provided it is defined
\be
\phi^{(i)}=\frac{1}{i!}\frac{d^{i}\phi^{(0)}}{d\Delta^{i}}.
\ee
It is then easy to see that
\be
\langle{\varphi^{(i)}}_{k}{\varphi'^{(j)}}_{k'}\rangle=\frac{1}{i!}
\frac{1}{j!}\frac{d^{i}}{d\Delta^{i}}\frac{d^{j}}{d\Delta'^{j}}
\langle{\varphi^{(0)}}_{k}{\varphi'^{(0)}}_{k'}\rangle . \ee The
corralator in the left--hand side is obtained from (38), and in
differentiating with respect to the weights one should regard the
constants $A_K$ as functions of the weights.

One can similarly define Jordanian blocks of full--fields. This
generalization is obvious:
\be
\phi^{(ij)}(z,\bar{z},\theta,\bar{\theta})=\frac{1}{i!}\frac{1}{j!}
\frac{d^{i}}{d\Delta^{i}}\frac{d^{j}}{d\Delta^{j}}\phi^{(0)}
(z,\bar{z},\theta,\bar{\theta}).
\ee
The correlators of the component--fields are then
\be
\langle\varphi^{(ij)}_{k\bar{k}}\varphi^{(lm)}_{k'\bar{k}'}\rangle=
\frac{1}{i!j!l!m!}\frac{d^{i}}{d\Delta^{i}}\frac{d^{j}}{d\bar{\Delta}^{j}}
\frac{d^{l}}{d\Delta'^{l}}\frac{d^{m}}{d\bar{\Delta}'^{m}}\langle
\varphi^{(00)}_{k\bar{k}}(z,\bar{z})\varphi'^{(00)}_{k'\bar{k}'}
(z',\bar{z}')\rangle . \ee The correlator in the right--hand side
is obtained from (40), and in differentiating with respect to the
weights, $A_{K\bar K}$'s are regarded as functions of the weights.


\begin{thebibliography}{99}
\bibitem{10} Joseph Polchinski; ``What is string theory'',
             hep--th/9411028
\end{thebibliography}
\end{document}